\documentclass[aps,prl,twocolumn, footinbib,longbibliography,superscriptaddress,preprintnumbers]{revtex4-2}
\usepackage{breakurl}
\usepackage{hyperref}
\usepackage{amsmath}
\usepackage{amssymb}
\usepackage{amsthm}
\usepackage{graphicx}
\usepackage{xcolor}

\renewcommand{\thanks}[1]{\footnote{#1}}
\newcommand{\bea}{\begin{eqnarray}}
\newcommand{\eea}{\end{eqnarray}}
\newcommand{\be}{\begin{eqnarray}}
\newcommand{\ee}{\end{eqnarray}}

\def\ie{\begin{equation}\begin{aligned}}
\def\fe{\end{aligned}\end{equation}}

\def\ie{\begin{equation}\begin{aligned}}
\def\fe{\end{aligned}\end{equation}}

\def\cO{{\cal O}}
\def\cP{{\cal P}}

\def\cbO{{\bf{\cal O}}}

\def\Tr{{\rm Tr}}
\def\det{{\rm det \,}}

\begin{document}
\preprint{QMUL-PH-23-13}
\title{Integrated correlators in $\mathcal{N}=4$ SYM beyond localisation}
\author{Augustus Brown}
\affiliation{Centre for Theoretical Physics, Department of Physics and Astronomy, Queen Mary University of London, London, E1 4NS, UK}
\author{Paul Heslop}
\affiliation{Centre for Particle Theory \& Department of Mathematical Sciences,  Durham University, Durham DH1 3LE, UK}
  \author{Congkao Wen}
\affiliation{Centre for Theoretical Physics, Department of Physics and Astronomy, Queen Mary University of London, London, E1 4NS, UK}
  \author{Haitian Xie}
\affiliation{Centre for Theoretical Physics, Department of Physics and Astronomy, Queen Mary University of London, London, E1 4NS, UK}
\begin{abstract}
We study integrated correlators of four superconformal primaries $\mathcal{O}_{p}$ with arbitrary charges $p$ in $\mathcal{N}{=}4$ super Yang-Mills theory (SYM). 
The  
$ \langle \mathcal{O}_{2} \mathcal{O}_{2} \mathcal{O}_{p} \mathcal{O}_{p} \rangle$ integrated correlators can be computed by supersymmetric localisation, whereas correlators with more general charges are currently not accessible from this method and in general contain complicated multi-zeta values. Nevertheless we observe that if one sums over the contributions from all different channels in a given correlator, then all the multi-zeta values (and products of zeta's) cancel leaving only $\zeta(2\ell{+}1)$ at $\ell$-loops.  We then propose an exact expression of such integrated correlators  in the planar limit, valid for arbitrary 't Hooft coupling. The expression matches with the known exact localisation-based  results for specific charges,  as well as with all existing perturbative and strong-coupling results in the literature for more general charges. As an application, our result is used to determine  certain $7$-loop Feynman integral periods and fix previously unknown coefficients in the correlators at strong coupling.

\end{abstract}

\maketitle


\section{Introduction}

Correlation functions of four superconformal primary operators, $\langle p_1\, p_2\, p_3\, p_4 \rangle :=\langle \mathcal{O}_{p_1}  \mathcal{O}_{p_2} \mathcal{O}_{p_3} \mathcal{O}_{p_4} \rangle$, in  $\mathcal{N}=4$ super Yang-Mills (SYM) are quantities of great interest, which have been computed perturbatively in the planar theory to ten loops in the 't Hooft coupling $\lambda$~\cite{Eden:2012tu,Bourjaily:2016evz,Caron-Huot:2021usw} as well as at strong coupling to $O(\lambda^{-3})$~\cite{Goncalves:2014ffa,Rastelli:2016nze,Alday:2017xua,Binder:2019jwn,Drummond:2019odu,Drummond:2020dwr,Abl:2020dbx,Aprile:2020mus,Alday:2023mvu}
 (see the review~\cite{Heslop:2022xgp} containing further references).
Here $\mathcal{O}_{p}$ is a superconformal primary  single-particle $1/2$-BPS operator with charge $p$ 
\ie
\label{Opdef}
	\cO_p(x,y)  = \Tr(\phi(x,y)^p)+\ldots , 
 \fe
where the dots denote multi-trace terms and $\phi(x,y)$ are the six fundamental scalars, with $x$ the spacetime variable and $y$ an internal equivalent variable packaging together the six scalars into a single object.
There are not many quantities in quantum field theories with more than two dimensions that are known exactly.
However, reference \cite{Binder:2019jwn} showed that a special class of such four-point functions, $ \langle 22pp\rangle$, may be computed exactly in both the Yang-Mills coupling $\tau$ and number of colours $N$, when they are integrated over their spacetime dependence with a certain choice of measure (equivalent to integrating over all four spacetime points modulo the conformal group~\cite{Wen:2022oky}). This remarkable fact arises by relating the integrated correlators to the  partition function of $\mathcal{N}=2^*$ SYM on the four-sphere that can be computed using localisation~\cite{Pestun:2007rz, Nekrasov:2002qd}\footnote{See also \cite{Drukker:2000rr, Pestun:2007rz} and more recently \cite{Pufu:2023vwo} for exact results   in $\mathcal{N}=4$ SYM using localisation involving a Wilson line.}. Exact finite-$N$ and finite-$\tau$ expressions are indeed obtained by exploring the localisation formula, see \cite{Dorigoni:2021guq, Dorigoni:2022zcr} for $\langle 2222\rangle$ and \cite{Paul:2022piq, Paul:2023rka, largep} for more general $\langle 22pp\rangle$ correlators. The supersymmetric localisation technique however does not extend to general correlators beyond the $\langle 22pp \rangle$ case~\cite{Binder:2019jwn,Gerchkovitz:2016gxx}\footnote{This is because, to preserve supersymmetry, the higher-charge operators $\mathcal{O}_p$ can only be inserted at northern and southern poles of the four-sphere and thus one can insert at most two such higher-charge operators.}, which we will study in this letter.

We consider integrated correlators for the most general four-point functions  $\langle p_1\, p_2\, p_3\, p_4 \rangle$. In particular, we observe that a great simplification occurs at five loops if we sum over all contributions from different $SU(4)_R$ channels contributing to a particular correlation function. Inspired by this as well as explicit perturbative results, we will present a remarkably simple and exact expression (valid for arbitrary $\lambda$) for any integrated correlator, summed over $SU(4)_R$ channels,  in the planar limit. Besides reproducing the $\langle 22pp\rangle$ case in~\cite{Binder:2019jwn}, the expression agrees with the existing results in the literature for more general correlators at both weak coupling \cite{Eden:2011we, Eden:2012tu} and strong coupling \cite{Abl:2020dbx, Aprile:2020mus}. Importantly, the strong coupling results had no input in obtaining the formula, and thus provide very strong support. 
  The strong coupling results will also be used to constrain unfixed parameters in the dual AdS amplitudes \cite{Abl:2020dbx, Aprile:2020mus}.

\section{Perturbative integrated correlators and periods}

We begin by considering integrated correlators perturbatively in the planar limit where all correlators are given by  a single object due to a ten-dimensional (10d) symmetry discovered in~\cite{Caron-Huot:2021usw} (see also \cite{Caron-Huot:2018kta,Caron-Huot:2022sdy}). In order to manifest this symmetry, it is useful to introduce a single operator that generates all the single-particle operators
\ie
\label{bcO}
\cbO(x,y) =  \sum_{p=2}^ \infty \frac1p \left(\frac{16\pi^4}c\right)^{p/4}\cO_p(x,y)\ ,
\fe
with $c=(N^2-1)/4$. Then (in the planar perturbation theory) the claim is that four-point functions of {\em all} single-particle $1/2$-BPS operators $\langle p_1p_2p_3p_4\rangle$
combine together into the following  {\em master correlator} 
\begin{align}
\label{master}
	&\!\!\!\!\!	\langle \cbO(x_1,y_1) \cbO(x_2,y_2) \cbO(x_3,y_3) \cbO(x_4,y_4) \rangle= \text{free  part} \\
  + &\, \frac{\mathcal{I}_4(x_i, y_j)}{2c}  \sum_{\ell=1}^\infty  \left(\! \frac{\lambda}{4\pi^2} \! \right)^\ell  {1\over \ell!} \! \int \! \frac{d^4x_5}{(-4\pi^2)}\ldots \frac{d^4x_{4+\ell}}{(-4\pi^2)}f^{(\ell)}({\bf x}_{ij}^2)\, ,  \nonumber
\end{align}
where $\lambda = N g^2_{_{Y\!M}}$ is the 't Hooft coupling, and $\mathcal{I}_4(x_i, y_j)$ is a known prefactor arising from superconformal symmetry. The bold spacetime invariants ${\bf x}_{ij}^2$ combine the external and internal variables into a natural ten dimensional object
\begin{align}\label{10dx}
	{\bf x}_{ij}^2:=x_{ij}^2 -y_{ij}^2=x_{ij}^2(1-g_{ij})\, ,
\end{align}
with $g_{ij}:=y_{ij}^2/x_{ij}^2$. 
Note however that only the external variables have non-trivial $y$'s, so $y_i=0$ if $i>4$, and thus $g_{ij}=0$ if $i$ or $j>4$. 
The function  $f^{(\ell)}({\bf x}_{ij}^2)$ is given as a sum over planar $f$-graphs~\cite{Eden:2011we,Eden:2012tu} (see also \cite{Bourjaily:2015bpz, Bourjaily:2016evz})
\begin{align}
	f^{(\ell)}({\bf x}_{ij}^2) &= \sum_{\alpha} c_\alpha^{(\ell)} f^{(\ell)}_\alpha({\bf x}_{ij}^2) \, , \\
	f^{(\ell)}_\alpha({\bf x}_{ij}^2)&= \frac 1 {|\text{aut}(\alpha)|} \sum_{\sigma \in S_{\ell+4}} \prod_{i,j=1}^{4+\ell}\frac{1}{ ({\bf x}_{\sigma_i\sigma_j}^2)^{e^\alpha_{ij}}} \, ,
\label{eq:f-graph}
\end{align}
where $\alpha$ are $(4+\ell)$-point graphs with net weight 4 at each vertex, $e^\alpha_{ij}$ is the number of edges between point $i$ and $j$ (we choose  a particular labelling of the graph to define $e^\alpha_{ij}$; numerator edges between $i$ and $j$ are negative). Here we sum over full permutation symmetry $S_{\ell+4}$~\cite{Eden:2011we,Eden:2012tu}, and ${|\text{aut}(\alpha)|}$ is the symmetry factor of the graph.

We now consider the integrated correlators. These have been defined (in the $\langle 22pp \rangle$ cases) by integrating the correlators (omitting the `free part' and divided by the prefactor $\mathcal{I}_4(x_i, y_j)$ as well as the factor $g_{34}^{p-2}$) over a certain measure. As shown in~\cite{Wen:2022oky}, this measure is equivalent to simply integrating over the four external variables (modulo the conformal group)
\begin{equation} \label{eq:measure}
	\int d\mu \ldots = -\frac{1}{\pi^2} \int\frac{ d^4x_1\dots  d^4x_4}{\text{vol}(SO(2,4))} \ldots \ .
\end{equation}
Note that it only makes sense to integrate in this way conformally covariant objects with conformal weight 4 at each of the four points $x_i$ against this measure.  Dividing by the factor $g_{34}^{p-2}$ gives the correct weight to be integrated thus to define the integrated $\langle 22pp\rangle$ correlator. 

This also shows us the natural way to define integrated correlators of arbitrary charges $\langle p_1p_2p_3p_4 \rangle$. These are more general polynomials of $g_{ij}$ (i.e. unlike $\langle 22pp\rangle$, they have multiple $SU(4)_R$ channels), so we can't simply divide by the $g_{ij}$ factors, but we instead integrate separately each coefficient of the $g_{ij}$ polynomial. Equivalently, we write the correlator as a function of $x^2_{ij},g_{ij}$ rather than $x^2_{ij},y^2_{ij}$ and treat $g_{ij}$ as constants.


Combined with the 10d symmetry \eqref{master} then all such integrated correlators are obtained from the {\em integrated master correlator} 
\begin{align}
\label{eq:def-general} 
\!\!\! \mathcal{C}(\lambda; g_{ij}) \!  := \!-\! \sum_{\ell=1}^{\infty} \left(\! \frac{\lambda}{4\pi^2} \! \right)^\ell \!\! \int\! \frac{d^4x_1 \dots d^4x_{4+\ell}}{\text{vol}(SO(2,4))}	\frac{f^{(\ell)}\big(x_{ij}^2(1 {-}g_{ij})\big) }{\pi^2 \ell! (-4\pi^2)^\ell} \, .
\end{align}
Here we integrate over {\em all} $(4+\ell)$ (internal and external) points. For every individual term in the permutation sum in a particular $f$-graph contribution 
$f^{(\ell)}_\alpha$~\eqref{eq:f-graph}, the $x_{ij}$ and $(1{-}g_{ij})$ terms completely factorise and after integration, the $x_{ij}$ contributions from a given $f$-graph contribute equally, all giving the period of the graph $f_\alpha^{(\ell)}$.
Thus all higher-charge integrated correlators package together as 
\begin{align} \label{eq:f-gij}
\!\!\! \mathcal{C}(\lambda; g_{ij})
=& -\sum_{\ell=1}^{\infty} \left(\! \frac{\lambda}{4\pi^2}\! \right)^\ell \frac{1}{\ell! (-4)^{\ell+1}} \sum_\alpha c_\alpha^{(\ell)} \cP{}\!\!_{f_\alpha^{(\ell)}} f^{(\ell)}_\alpha(1{-}g_{ij})\ ,
\end{align}
where the period is defined as
\ie \label{eq:period}
\cP{}\!\!_{f_\alpha^{(\ell)}} = {1\over (\pi^2)^{\ell+1}}  \int \frac{d^4x_1 \dots d^4x_{4+\ell}}{\text{vol}(SO(2,4))}	f^{(\ell)}_\alpha({x}_{ij}^2) \, .
\fe
Periods of Feynman integrals of the form \eqref{eq:period} have been studied quite extensively in the literature  \cite{Broadhurst:1995km, Schnetz:2008mp, Brown:2009ta, Schnetz:2013hqa,  Panzer:2016snt, HyperlogProcedures, Panzer:2014caa, Borinsky:2020rqs}, and these results have been utilised in the case of $\langle 2222 \rangle$~\cite{Wen:2022oky}.
 
Let us consider some examples. At one loop, we have
 \begin{align}
 	f^{(1)}(x_{ij}^2)= \frac1{\prod_{1\leq i<j \leq 5} x_{ij}^2 } \, ,
 \end{align}
 and so the integrated master correlator is simply  (recall $y_5=0$ and $g_{i5}=0$)
 \begin{align}
 	-{1\over 1! (-4)^1} \frac{\cP{}\!\!_{f^{(1)}}}{\prod_{ 1 \leq i<j \leq 4} (1-g_{ij}) }\ ,
 \end{align}
 with $\cP{}\!\!_{f^{(1)}} = 6 \zeta(3)$~\cite{Belokurov:1983km, Broadhurst:1995km}. 
 The correlators $\langle 22pp\rangle$ are extracted from this by taking the coefficient of $g_{34}^{p-2}$ and setting all $g$'s to zero. In this case they are all equal, in agreement with~\cite{Binder:2019jwn, Paul:2022piq}. 
At two loops, we have
\begin{align}
	f^{(2)}(x_{ij}^2)= \frac{1}{48}\sum_{\sigma \in S_6} \frac {x^2_{\sigma_1 \sigma_2}x^2_{\sigma_3 \sigma_4}x^2_{\sigma_5 \sigma_6}}{\prod_{1\leq i<j \leq 6} x_{ij}^2 } \, ,
\end{align}
and  the integrated master correlator becomes 
\begin{align}
\!\!\!	- {\cP{}\!\!_{f^{(2)}} \over 2! (-4)^2}\frac{ g_{12} g_{34} {+}  g_{13} g_{24} {+}g_{14}g_{23}  {-} 3\sum_{1 \leq i<j \leq 4}  g_{ij} {+}15 }
	{ \prod_{ 1 \leq i<j \leq 4} (1-g_{ij})  }\ , \nonumber
\end{align}
where $\cP{}\!\!_{f^{(2)}}=  20 \zeta(5)$~\cite{Belokurov:1983km, Broadhurst:1995km}. For $\langle 22pp\rangle$, after we set $g_{ij}$ to zero except $g_{34}$, it gives
\begin{align}
- {\cP{}\!\!_{f^{(2)}} \over 2! (-4)^2} \left(\frac{12}{1-g_{34}}+3\right)\ .
\end{align}
Expanding in $g_{34}$ then yields the  $\langle 22pp\rangle$ integrated correlators at two loops:  as $-{75 \zeta (5)}/{8}$ for $\langle 2222 \rangle$ and  $-{15 \zeta (5)}/{2}$ for all others, again 
 agreeing with~\cite{Binder:2019jwn, Paul:2022piq}.

 Proceeding similarly to higher loops, we obtain integrated correlators of all single-trace $1/2$-BPS operators in terms of the $f$-graph periods at higher loops.
 When specifying to the $\langle 22pp \rangle$ case, the resulting expressions all agree with the known results~\cite{Binder:2019jwn, Paul:2022piq}.

\section{Simplification and all order expression}
Crucially, starting from five loops we find multi-zeta values (and products of zeta's) on integrating higher-charge correlators. For example, for $\langle 4444 \rangle$ at five loops, we find, 
\ie \label{eq:4444}
& \hspace{-1.6cm} \frac{g_{12}^2 g_{34}^2}{40320}  \Big[7560 \zeta(5,3,3){+}14685615 \zeta (11){+}56700 \pi ^2 \zeta (9)
	\cr  
	+ \, 252 \pi ^4 \zeta (7)&{+}31500 \zeta (5)^2{+}6300 \zeta
	(3)^2 \zeta (5)-20 \pi ^6 \zeta (5)\Big]   \cr
	\!\!\!\!\!\!\!\!\!\!\!\!\!-\,
	\frac{g_{12}g_{23}g_{34}g_{14}}{40320}  &  \Big[7560 \zeta(5,3,3){+}569205 \zeta (11){+}56700 \pi ^2 \zeta (9) \cr
	 + \, 252 \pi ^4 \zeta (7)&{+}31500 \zeta (5)^2{+}6300 \zeta
	(3)^2 \zeta (5) {-} 20 \pi ^6 \zeta (5)\Big]	 \\
	&\hspace{1.5cm} + \ \text{3 terms of crossing} \, .
\fe
However we note that, intriguingly, in the sum of the two terms in \eqref{eq:4444} (despite different $g_{ij}$ factors!) everything but $\zeta(11)$ cancels.

Remarkably this pattern continues at higher charges: for any correlator at five loops, if we sum over all contributions in this way, so that all $SU(4)_R$ channels contribute equally, then one is always left with a single $\zeta(11)$. 
 This sum over channels can be implemented automatically by  formally replacing $g_{ij} \rightarrow \gamma_i \gamma_j$.  With this replacement the different $g_{ij}$ factors in a given $\langle p_1p_2p_3p_4 \rangle$ correlator contribute equally and we can see the above simplification directly for all correlators at the level of the integrated master correlator. The $\gamma_i$ carry the information of the operator charges (so the integrated $\langle p_1p_2p_3p_4 \rangle$ correlator, summed over channels, arises as the $\gamma_1^{p_1-2}\gamma_2^{p_2-2}\gamma_3^{p_3-2}\gamma_4^{p_4-2}$ coefficient.
Note $g_{ij} \rightarrow \gamma_i \gamma_j$ is irrelevant for $\langle 22pp \rangle$,  since the correlator has only a single $SU(4)_R$ channel.\footnote{More precisely, the correlator $\langle 22pp \rangle$ depends on $g_{ij}$ as $g_{34}^{p-2}$, therefore the condition $g_{ij}=\gamma_i \gamma_j$ makes no difference in this case.}

We therefore observe the striking feature in the integrated correlators with $g_{ij}\rightarrow \gamma_i \gamma_j$ that all multi-zetas cancel. We then initiated a careful examination of the resulting perturbative results which we knew up to five loops for all charges $p_i$.
By comparing with all order results of the $\langle 22pp\rangle$ family, known from localisation\cite{Binder:2019jwn}, we were then able to guess new all order results, first for the $\langle 33pp\rangle$ family of correlators, and later for the $\langle 44pp\rangle$ as well as other cases (such as $\langle 23pp{+}1\rangle$, $\langle 24pp{+}2\rangle$), thus obtaining functions of $p$ and the coupling $\lambda$. Although these non-trivial functions are obtained based on observation from lower-loop expressions, we are confident of their validity due to the strong coupling matching as well as other checks, which will be discussed later.

Furthermore, on lifting these results to the master correlator and rewriting the resulting symmetric polynomial in $\gamma_1,\gamma_2,\gamma_3,\gamma_4$ in terms of Schur poynomials
we found further dramatic simplification. The result of these investigations is then a proposal for  the planar integrated master correlator \eqref{eq:def-general} (thus yielding all integrated correlators $\langle p_1p_2p_3p_4 \rangle$ for all $p_i$ and for all coupling $\lambda$, summed over channels) via the remarkably simple formula 
\ie \label{eq:master-In}
\!\!\! \mathcal{C}(\lambda; \gamma_i \gamma_j) =   \!\sum_{\ell=1}^{\infty}   \lambda ^{\ell} \sum^{\infty}_{\nu=2}\frac{4 (-1)^{\nu+\ell+1}  \Gamma
 \!  \left(\ell {+} \frac{3}{2}\right)^2 \! \zeta (2 \ell {+} 1)  }{ \pi ^{ 2\ell +1}  \Gamma (\ell {+}2{-}\nu) \Gamma (\ell {+} \nu {+}1)}  F_\nu(\gamma_i)  \, , 
   \fe
   where $\ell$ is the number of loops  and we have set $g_{ij}\rightarrow \gamma_i \gamma_j$.
   The factor $F_\nu(\gamma_i)$, which contains the information of the operator charges, is given by
\ie \label{eq:F-gn}
F_\nu(\gamma_i) ={ \mathcal{S}_{\nu-2,\nu-2,0,0} (\gamma_i) - \mathcal{S}_{\nu-2,\nu-2,1,1} (\gamma_i)  \over  \prod_{1\leq i <j \leq 4}  (1-\gamma_i \gamma_j ) } \, , 
\fe
where $\mathcal{S}_{\nu_i} (\gamma_i)$ are the Schur polynomials 
\begin{equation}
 \mathcal{S}_{\nu_1,\nu_2,\nu_3,\nu_4} (\gamma_i) = \frac{\det(\gamma_i^{4+\nu_j-j})_{i,j=1,2,3, 4}}{\prod_{1\leq i<j\leq 4}(\gamma_{i}-\gamma_j)}\ .
\end{equation} 
Some comments are in order to illuminate the expression of $F_\nu(\gamma_i)$. The structure of the denominator in $F_\nu(\gamma_i)$ is expected from \eqref{eq:f-gij} and \eqref{eq:f-graph}. The fact that the numerator is given by some symmetric polynomial in $\gamma$'s is also not surprising because of the symmetric integration measure, thus the integrated correlators should have a permutation symmetry. What is striking is the appearance of the Schur polynomials with very special partitions, and more importantly the remarkable simplicity of the overall formula which is hidden unless written in terms of Schur polynomials! 

The integrated $\langle p_1 \,  p_2 \, p_3 \,  p_4 \rangle $ correlator  (summed over channels)  is then extracted from~\eqref{eq:master-In} by simply taking the coefficient of $\gamma_1^{p_1-2}\gamma_2^{p_2-2}\gamma_3^{p_3-2}\gamma_4^{p_4-2}$, 
\ie \label{eq:gen-fun}
\mathcal{C}_{p_1p_2p_3p_4}(\lambda) := \mathcal{C}(\lambda; \gamma_i \gamma_j) \Big{\vert}_{\gamma_1^{p_1-2}\gamma_2^{p_2-2}\gamma_3^{p_3-2}\gamma_4^{p_4-2}}\, .
\fe
We remark that for correlators at a fixed loop order $l$ or for a fixed correlator of charges $p_i$,  the summation over $\nu$ in \eqref{eq:master-In} is naturally truncated: the $\Gamma(\ell {+}2{-}\nu)$ in the denominator means all terms with $\nu > l{+}2$ vanish; whereas Schur polynomials with $\nu > \max(p_i)$ will not contribute to the $\langle p_1p_2p_3p_4 \rangle$ correlator.  
It is worth noting that there are intriguing relations among seemingly different integrated correlators because of the special form of $F_\nu(\gamma_i)$. For example, the integrated correlator of $\langle p_1,p_1{+}q_1,p_2,p_2{+}q_2\rangle$ is equivalent to that of $\langle p_1', p_1'{+}q_2, p_2',p_2'{+}q_1\rangle$ if $2p_1{+}q_1=2p'_1 {+} q_2$ and $2p_2{+}q_2=2p'_2 {+} q_1$. 

 Finally, we comment on the 10d symmetry and its role in our construction. It is incredibly useful in order to write down the generating function, and thus quickly obtain results for arbitrary charge correlators which we then used to obtain~\eqref{eq:master-In}.  However our formula \eqref{eq:master-In} is not a simple consequence of the 10d symmetry, which is a symmetry only at the integrand level, destroyed upon integration.   
Furthermore, the 10d symmetry {\em is} broken at strong coupling, but as we will see the expression~\eqref{eq:master-In} correctly reproduces all arbitrary charge integrated correlators at strong coupling as well.

\section{Re-summed expression}

The proposed all-order expression \eqref{eq:master-In} can be re-summed through a modified Borel transform (see e.g.~\cite{Russo:2012kj,Hatsuda:2015owa}) by using the following integral identity
\ie
\zeta(n)\, \Gamma(n{+}1) = 2^{n-1} \!\int^{\infty}_0 dw \frac{w^{n}}{\sinh^2(w)} \, .
\fe
Replacing $\zeta(2\ell+1)$ in \eqref{eq:master-In} by its integral representation using the above identity and performing the resummation,  we obtain
\ie \label{eq:BessJ}
\!\!\! \mathcal{C}(\lambda; \gamma_i \gamma_j)  =  \! \int_0^{\infty} \!\! \frac{w \, dw }{\sinh^2(w)} \sum_{\nu=2}^{\infty}  \left( J_{\nu-1}(u)^2{-}J_\nu(u)^2\right) \! F_\nu(\gamma_i)  \, ,   
   \fe
where $u= \frac{w \sqrt{\lambda }}{\pi }$ and $J_\nu(u)$ are Bessel functions.   For $\langle 22pp\rangle$, \eqref{eq:BessJ} reduces to the known result of \cite{Binder:2019jwn}
\ie \label{eq:22pp-B}
\mathcal{C}_{2,2,p,p}(\lambda) =\int_0^{\infty}  \frac{ w\, dw }{\sinh^2(w) }\left(J_1(u)^2-J_p(u)^2 \right) \, ,
\fe
while for $\langle 33pp\rangle$, for instance, we find 
\begin{align}
    \label{eq:33pp-B}
\mathcal{C}_{3,3,p,p} & (\lambda) = \int_0^{\infty}   \frac{ w\, dw }{\sinh^2(w) }\Big( 3 J_1(u)^2+4 J_2(u)^2 \\
&\!\!\!\!\!\! +J_3(u)^2  -2 J_{p-1}(u)^2-4 J_p(u)^2-2 J_{p+1}(u)^2 \Big)\,. \nonumber
\end{align} 
Similar expressions can be obtained from \eqref{eq:BessJ} for all other integrated correlators. 

Importantly, the expression \eqref{eq:BessJ} is valid for arbitrary $\lambda$, allowing us to study the integrated correlators in the strong-coupling regime, which we will consider in the next section.

\section{Strong coupling}

The strong-coupling expansion of the integrated correlators can be obtained straightforwardly from \eqref{eq:BessJ} by using the Mellin-Barnes representation of products of Bessel functions~\cite{Binder:2019jwn}, and we find
\begin{align} \label{eq:large-lambda}
& \mathcal{C}(\lambda; \gamma_i \gamma_j) \Big{\vert}_{\rm strong} = \sum_{\nu=2}^{\infty}  \Big( \frac{1}{2\nu\left(\nu - 1 \right)}\, +
\\
&  \sum_{n=1}^{\infty} \frac{4 n (-1)^n \, \Gamma\!
   \left(n {+} \frac{1}{2}\right) \Gamma \! \left(\nu {+} n{-}\frac{1}{2}\right)\zeta(2 n {+} 1)}{\lambda^{n +\frac{1}{2}} \sqrt{\pi} \Gamma (n)
   \Gamma\left(\nu {-} n {+} \frac{1}{2} \right)}   \Big) \,  F_\nu(\gamma_i)  \, . \nonumber
   \end{align}
This can then be compared with known results for four-point correlators of general charges at strong coupling. The string corrections are fully determined by a simple effective action of a single massless 10d scalar on AdS${}_5\times $S$^5$\cite{Abl:2020dbx}
and are completely fixed up to order $\lambda^{-5/2}$ and $\lambda^{-3}$~\cite{Drummond:2020dwr,Abl:2020dbx, Aprile:2020mus} under a certain $\mathbb{Z}_2$-symmetry assumption\footnote{The assumption is that coefficients in the effective action on AdS${}_5\times$ S${}^5$ have an expansion in $\alpha'^2/R^4$ rather than $\alpha'/R^2$, so that odd powers in $\alpha'/R^2$ vanish, which looks natural when the effective action is viewed as arising from a general curved space action. See \cite{Abl:2020dbx}}. 
Note that this 10d effective action is very different from the 10d symmetry of the {\em integrand} of the perturbative correlators. The tree-level supergravity result possesses 10d conformal symmetry at the level of the function itself (not just the integrand), but the higher $\alpha'$ corrections break this. Instead they can be written as a 10d integral over AdS${}_5 \times S^5$. The key point for us is simply that this gives us access to the full master correlator at strong coupling, generating correlators of all charges.
One can straightforwardly integrate the space-time dependence of these strong-coupling results with the measure~\eqref{eq:measure} analytically, directly from their  Mellin space expressions, by using the techniques outlined in appendix B of~\cite{Chester:2020dja}. 
These give highly non trivial symmetric functions of the four $\gamma_i$ at each order.
Nevertheless we find complete agreement for all the correlators, for all $p_i$  at 
supergravity level, $\lambda^0$, as well as at orders $\lambda^{-3/2}$, $\lambda^{-5/2}$ and $\lambda^{-3}$  (where it adds further confirmation to  the $\mathbb{Z}_2$-symmetry).

At order $\lambda^{-7/2}$ there remain unfixed coefficients but our result~\eqref{eq:large-lambda} is both consistent with these and indeed constrains them further. 
There are 11 coefficients occurring as possible non-equivalent terms in the effective action,  and comparison with our integrated correlator result~\eqref{eq:large-lambda} fixes five of these. If in addition we  compare with the recently available $\langle 2222 \rangle$ correlator that has been completely fixed at this order~\cite{Alday:2023mvu}, we further fix three coefficients. The $\mathbb{Z}_2$ symmetry then fixes two more coefficients to zero and we are left with a single unfixed coefficient. Explicitly in the notation of~\cite{Abl:2020dbx} we fix the nine coefficients:
\begin{align}
  \begin{array}{lll}
\!\!\! A_4=-\frac{1575 \zeta (7)}{4}\, , & C_2=\frac{641 \zeta (7)}{16}\, , & D_1=0\, , \\
\!\!\! E_1=0 \, , & F_0=\frac{\zeta (7)}{2}\,, & G_{1;0}=-\frac{11 \zeta (7)}{64} \, ,\\
\!\!\! G_{2;0}=\frac{71 \zeta (7)}{64} \, ,& G_{3;0}=\frac{141 \zeta (7)}{256}\, , & G_{5;0}=-\frac{51
   \zeta (7)}{64} \, , 
\end{array}
\end{align}
together with a relation  between the remaining two, $ B_2=20 G_{4;0} +\frac{259 \zeta (7)}{4}$. Finally in~\cite{Aprile:2020mus} the authors found one additional constraint arising from the OPE, which goes  beyond what the effective action predicts (which are there referred to as the `rank constraints').  This additional constraint is then enough to fix the remaining free parameter, giving $B_2=\frac{170839 \zeta (7)}{1664}$ and $G_{4;0}=\frac{12619 \zeta (7)}{6656}$.


It is important to note that, unlike the small-$\lambda$ expansion \eqref{eq:master-In}, the large-$\lambda$ expansion \eqref{eq:large-lambda} is asymptotic and not Borel summable. The asymptotic series should be completed with exponentially decayed terms, which can be obtained by the means of resurgence~\cite{Dorigoni:2014hea}. By performing  resurgent analysis following~\cite{Dorigoni:2021guq, Hatsuda:2022enx}, we find
   \begin{align}
  \label{eq:exp}
 &  \Delta  \mathcal{C}(\lambda; \gamma_i \gamma_j) = \pm  {i\over 2}   \sum_{\nu=2}^{\infty} (-1)^\nu (2 \nu{-}1)^2 \Big(\frac{8 \,\text{Li}_0(z)}{  (2 \nu{-}1)^2  }  \\
   & + \, \frac{2 \ \text{Li}_1(z)}{\lambda^{1/2}}  +\frac{(4 \nu^2{-}4 \nu{+}5) \text{Li}_2(z)}{4\ \lambda }+ \ldots \Big) \,  F_\nu(\gamma_i)  \, , \nonumber    
   \end{align}
   where $z= e^{-2 \sqrt{\lambda }}$. Holographically, these terms behave as $e^{-2L^2/\alpha'}$ ($L$ is the AdS length scale), which indicates they may arise from world-sheet instantons. 

\section{Modular invariance}

   The S-duality of $\mathcal{N}=4$ SYM \cite{Montonen:1977sn, Goddard:1976qe} implies that the correlators of the operators we consider here should be $SL(2,\mathbb{Z})$ invariant. Therefore, the large-$N$ expansion of the integrated correlators with {\em fixed coupling $\tau$} should be expressed in terms of modular functions. More explicitly, as in the cases of $\langle 2222\rangle$ \cite{Chester:2019jas, Dorigoni:2021guq, Dorigoni:2022cua} (and more generally $\langle 22pp\rangle$ \cite{Paul:2022piq,Paul:2023rka, largep}), we expect the power series terms \eqref{eq:large-lambda} are replaced by the non-holomorphic Eisenstein series $E(s; \tau, \bar \tau)$, whereas the exponentially decayed terms \eqref{eq:exp} should be expressed in terms of $D_N(s; \tau, \bar \tau)$ introduced in \cite{Dorigoni:2022cua} (see also \cite{Luo:2022tqy} in a different context). These modular functions are defined as 
   \begin{align}
   E(s; \tau, \bar \tau) &= \sum_{(m,n) \neq (0,0)}  {\tau_2^s \over \pi^s |m+n\tau|^{2s}} \, , \\
   D_N(s; \tau, \bar \tau) &= \sum_{(m,n) \neq (0,0)} e^{-4\sqrt{N \pi } \frac{|m+n\tau|}{\sqrt{\tau_2}}} {\tau_2^s \over \pi^s |m+n\tau|^{2s}} \, , \nonumber
   \end{align}
   where $\tau:= \theta/(2\pi) + i 4\pi/g_{_{Y\!M}}^2 =\tau_1 + i \tau_2$. 
   The planar results \eqref{eq:large-lambda} and \eqref{eq:exp} then allow us to determine the first few orders in the large-$N$ (fixed $\tau$) expansion, 
   \ie \label{eq:S-duality}
 & \!\!\!\!  \mathcal{C}(\tau, \bar{\tau}; \gamma_i \gamma_j) =      \sum_{\nu=2}^{\infty}  \Big[ \frac{1}{2 (\nu {-} 1) \nu} - \frac{2 \nu{-}1 }{2^4 N^{3 \over 2}}E(3/2; \tau, \bar{\tau}) 
\cr
&~~~~~~~ +\frac{3 (2 \nu{-}3) (4 \nu^2{-}1)
}{2^8 N^{5\over 2}}     E(5/2; \tau, \bar{\tau}) + \ldots 
\cr & ~~~~~~~
\pm 2i (-1)^\nu D_N(0; \tau, \bar \tau) + \ldots \Big] \,  F_\nu(\gamma_i) \, , 
   \fe
where ``dots" denote higher-order terms \footnote{To determine the higher-order terms, one would need results beyond the planar limit. For instance, we expect the next order is given by a sum of $E(3/2; \tau, \bar{\tau})$ and $E(7/2; \tau, \bar{\tau})$; the coefficient of $E(7/2; \tau, \bar{\tau})$ is fixed by our planar result, but not $E(3/2; \tau, \bar{\tau})$.}. The expression $\mathcal{C}(\tau, \bar{\tau}; \gamma_i \gamma_j)$ agrees with known results in the literature when there is an overlap in charges and to the order that we can determine. The fixed-$\tau$ result is beyond the 't Hooft limit; in particular, it contains non-perturbative Yang-Mills instanton contributions.

\section{New periods and the 10d lightlike limit}

Finally we return to the perturbative regime where the all-order expression~\eqref{eq:master-In} can be compared directly with the results obtained from four-point functions integrated in terms of periods~\eqref{eq:f-gij}.  
At six loops, there are $26$ periods and the integrands can be found in~\cite{Eden:2012tu}. They are highly non-trivial $7$-loop Feynman integral periods \footnote{There is one additional loop compared to the correlator because of the integration over the external points.}. We have evaluated $16$ of them explicitly using {\tt HyperlogProcedures}~\cite{HyperlogProcedures}, and we find the proposed expression \eqref{eq:master-In} is perfectly consistent with these Feynman integral results. Furthermore \eqref{eq:master-In} allows us to then determine the remaining unknown periods. We find \eqref{eq:master-In} fixes all the periods except a single one, which can be further determined by exploiting a fascinating connection~\cite{Caron-Huot:2021usw} between the master correlator and the {\em octagon}, ${\mathbb O}$, introduced in~\cite{Coronado:2018ypq,Coronado:2018cxj}, as we will discuss now.  

 As proposed in~\cite{Caron-Huot:2021usw}, in a 10d lightlike limit ${\bf{x}}_{i,i+1}^2=x_{i,i+1}^2-y_{i,i+1}^2 \rightarrow 0$, the master correlator~\eqref{master} reduces to the octagon, ${\mathbb O}$ (squared), 
 \begin{align}\label{ll}
 \lim_{{\bf{x}}_{i,i+1}^2\rightarrow 0}  \frac{\!\!\!\!\langle \cbO \cbO\cbO \cbO \rangle}{\langle \cbO \cbO\cbO \cbO \rangle_\text{free}\!\!\!\!}
 =M^2  \, ,
 \end{align}
where $M{=}\,{{\mathbb O}}/{{\mathbb O}_{\text{free}}}$
can be viewed as a massive four-particle amplitude on the Coulomb branch~\cite{Caron-Huot:2021usw}. This relation~\eqref{ll} is valid at the level of perturbative integrands, but both sides are finite and can be integrated. The octagon is given (using integrability) in terms of sums of products of known ladder integrals~\cite{Belitsky:2020qir}, whereas the correlator is given in terms of non-trivial four-point conformal integrals. This relation then predicts non-trivial relations between these conformal integrals, which has been previously confirmed (numerically) to four loops~\cite{Caron-Huot:2021usw}.  

Now the 10d lightlike limit together with $g_{ij} \rightarrow \gamma_i \gamma_j$ implies $\gamma_1= \gamma_3 = 1/\gamma_2= 1/\gamma_4:= \gamma$. Making this substitution as well as integrating over both sides of \eqref{ll} with the measure \eqref{eq:measure} allows a direct all orders comparison. For the left hand we use our exact formula  \eqref{eq:master-In}. The right hand side arising from the octagon~\cite{Caron-Huot:2021usw},  is also completely  known to all orders in this limit~\footnote{After the substitution, $\gamma_1= \gamma_3 = 1/\gamma_2= 1/\gamma_4:= \gamma$, only single ladders and the product of two single ladders  contribute to $(M)^2$, both of which are known to all orders. Indeed the period of the product of two ladders equals the period of the single ladder at the same total loop order $\ell$, $2\binom{2 \ell+1}{ \ell+1}\zeta(2 \ell+1)$~\cite{Dorigoni:2021guq}.}.  
We find indeed both sides of~\eqref{ll} lead to the same result
\begin{align}
-\sum_{\ell} \left({ \lambda \over 4\pi^2} \right)^{\ell}  {\ell+1 \over 2^{2 \ell+1}}
 \binom{2
   \ell{+}2}{\ell{+}1}  \frac{ (\gamma^2-1)^{2
   \ell} }{\gamma^{2\ell} }
   \zeta (2 \ell+1) \, . 
   \end{align}
   
We have further verified the relation~\eqref{ll} up to six loops  (again integrating over external points) without imposing $g_{ij} \rightarrow \gamma_i \gamma_j$, using the results of Feynman integral periods determined from \eqref{eq:master-In}. All these simultaneously provide further consistency checks of both our proposal \eqref{eq:master-In} and the relation \eqref{ll}.  
In the process, we also confirm many integral relations  predicted from it, at least at the level of further integration over external points. As an application, these integral relations express the aforementioned unfixed period in terms of certain (integrated) triple products of ladders, which we are able to compute using {\tt HyperlogProcedures} unlike the original period. More detailed discussion of the integrated correlators at six loops and the evaluation of the periods at this order can be found in the appendix \ref{app:6loops} as well as the attached ancillary {\tt Mathematica} notebook.

\section{Conclusion}
We have presented an exact formula for the integrated correlators of arbitrary charges in the planar limit. We should emphasise that the formula was obtained purely based on examining data to five loops in perturbation theory $O(\lambda^5)$ (but for all correlators $\langle p_1p_2p_3p_4\rangle$). The fact that the formula is then consistent with the periods entering six loops $O(\lambda^6)$ computed directly, as well as with the octagon, and furthermore agrees with all strong coupling results to $\lambda^{-7/2}$  for correlators of all charges gives us very high confidence in our proposal.  

The result reveals a remarkable simplicity of these general integrated correlators at both weak and strong coupling, even though they are currently inaccessible by supersymmetric localisation methods.  A natural question that arises then is what the origin of this simplicity is, and in particular what is the meaning of, or reason for, the $g_{ij}\rightarrow \gamma_i \gamma_j$ replacement. A related question is if there are some other operations in $g_{ij}$ space one could perform, which still yield simple and interesting results.

Another important topic for further investigation concerns the extension of our results to the non-planar sectors and even finite $N$. It has been shown that, for the special case $\langle 22pp\rangle$, the integrated correlators obey Laplace-difference equations that relate them with different $N$ and charges~\cite{Dorigoni:2021bvj, Dorigoni:2021guq, Paul:2022piq, recp}. It would be fascinating if the more general integrated correlators studied in this letter obey similar relations. Seeing any such structures requires beyond the planar limit. 

We would also like to explore the integrated correlators with a different integration measure~\cite{Chester:2020dja, Chester:2020vyz}, which was originally introduced for the correlator $\langle 2222 \rangle$.  In perturbation theory, they can again be understood as periods of Feynman integrals~\cite{Wen:2022oky}. 
We leave a systematic study of the second integrated correlators as a future research direction.

\section*{acknowledgments}
{\small 
 We would like to thank Francesco Aprile,  Shai Chester, Frank Coronado, Daniele Dorigoni, Alessandro Georgoudis, 
 Tobias Hansen, Arthur Lipstein, and Yifan Wang for helpful discussions.
 CW and HX are supported by Royal Society University Research Fellowships, UF160350 and URF\textbackslash R\textbackslash 221015, and AB is supported by a Royal Society funding  
 RF\textbackslash ERE\textbackslash 210067. }

\appendix 

\setcounter{secnumdepth}{2}

\section{Explicit results of $7$-loop periods and relations to ladder integrals} \label{app:6loops}

Here we present some examples of $7$-loop periods (they enter in the integrated correlators at six loops) that we determined using our exact expression of the integrated correlators as well as the connection between the correlators and the octagon. The complete results (and the corresponding integrands) of all the 26 periods that are relevant at this order can be found in the attached ancillary {\tt Mathematica} notebook. 

Below is an example that we have difficulty to evaluate using {\tt HyperlogProcedures}, 
\begin{align}
& \!\! \cP{}\!\!_{f_2^{(6)}} \!=\! {1\over (\pi^2)^{7}} \! \int \! \frac{d^4x_1 \dots d^4x_{10}}{\text{vol}(SO(2,4))} \frac{x_{18}^2 x_{2\,10}^2 x_{46}^2 x_{47}^2}{x_{12}^2 x_{13}^2 x_{14}^2 x_{15}^2
   x_{16}^2 x_{23}^2 x_{26}^2 x_{27}^2 x_{28}^2 
   } 
   \cr
   & \frac{1}{x_{34}^2 x_{38}^2 x_{45}^2 x_{48}^2 x_{49}^2  x_{4\,10}^2 x_{56}^2  x_{5\,10}^2 x_{67}^2 x_{6\,10}^2 x_{78}^2
   x_{79}^2 x_{7\,10}^2 x_{89}^2 x_{9\,  10}^2} \, . \nonumber
\end{align}
We see that $\cP{}\!\!_{f_2^{(6)}}$ is an extremely non-trivial $7$-loop Feynman integral (one may set, e.g. $(x_1,x_2,x_3) \rightarrow (0,1,\infty)$ using conformal symmetry). 
However, we find that $\cP{}\!\!_{f_2^{(6)}}$ can be fixed by using our exact result \eqref{eq:master-In}, which leads to
\begin{align}
\cP{}\!\!_{f_2^{(6)}} &=   2880 \zeta (5) \zeta (5,3)+1440 \zeta (5,5,3)-720 \zeta (7,3,3)\cr 
  &\,  -106632 \zeta (13) + 8580
   \pi ^2 \zeta (11)+176 \pi ^4 \zeta (9)\cr 
  &\,+2800 \zeta (5) \zeta (7)+1680 \zeta (3)^2 \zeta
   (7)+\frac{64 \pi ^6 \zeta (7)}{21}\, .
\end{align} 
In a similar fashion, \eqref{eq:master-In} fixes all the 25 of 26 periods, except the following period
\begin{align}
& \!\!\!\!  \cP{}\!\!_{f_1^{(6)}} \!=\!    {1\over (\pi^2)^{7}} \! \int \! \frac{d^4x_1 \dots d^4x_{10}}{\text{vol}(SO(2,4))} \frac{x_{19}^2 x_{2\,10}^2 x_{36}^2 x_{58}^2}{x_{12}^2 x_{13}^2 x_{14}^2 x_{15}^2
   x_{16}^2 x_{23}^2 x_{26}^2 x_{27}^2 x_{28}^2  } \cr 
 &  \frac{1}{x_{34}^2 x_{38}^2 x_{39}^2
   x_{45}^2 x_{49}^2 x_{56}^2 x_{59}^2 x_{5\,10}^2 x_{67}^2 x_{6\,10}^2 x_{78}^2
   x_{7\,10}^2 x_{89}^2 x_{8\,10}^2 x_{9\,10}^2} \, . \nonumber
\end{align}
To be precise, there are three other periods   also not determined using \eqref{eq:master-In}, but they are all linearly related to $\cP{}\!\!_{f_1^{(6)}}$, so that in obtaining \eqref{eq:master-In}, $\cP{}\!\!_{f_1^{(6)}}$ cancels out. 

As discussed in the main text, $\cP{}\!\!_{f_1^{(6)}}$ can be shown to be related to products of ladder integrals using the connection between the correlators and octagon \eqref{ll}, which leads to
\begin{align}
& \!\!\!\!   \cP{}\!\!_{f_1^{(6)}} = 6 L_{1,1,4}-L_{1,2,3}-6880 \zeta (5)
   \zeta (5,3)+1536 \zeta (5,3,3)\cr
   & -608 \zeta
   (5,5,3)-640 \zeta (7,3,3)+408252 \zeta (13)-35640
   \pi ^2 \zeta (11) \cr
   & -124064 \zeta (11)-5760 \zeta (3)
   \zeta (9)-368 \pi ^4 \zeta (9)+11520 \pi ^2 \zeta
   (9) \cr
   & -13120 \zeta (5) \zeta (7)-4480 \zeta (3)^2
   \zeta (7)+\frac{512 \pi ^6 \zeta
   (7)}{189}+\frac{256 \pi ^4 \zeta (7)}{5} \cr
   & -7200 \zeta
   (3) \zeta (5)^2+3200 \zeta (5)^2+1280 \zeta (3)^2
   \zeta (5)-\frac{256 \pi ^6 \zeta (5)}{63}\, , \nonumber
\end{align}
where $L_{\ell_1, \ell_2, \ldots, \ell_n}$ are products of $n$ ladder integrals with $\ell_1, \ell_2, \ldots, \ell_n$ loops, integrated with the measure \eqref{eq:measure}. Now it is straightforward to evaluate $L_{1,1,4}$ and $L_{1,2,3}$ using {\tt HyperlogProcedures}, we then obtain
\begin{align}
 & \! \cP{}\!\!_{f_1^{(6)}} \!=\! -1760 \zeta (5) \zeta (5,3)+768 \zeta (5,3,3)-1696 \zeta (5,5,3) \cr 
  &+1120 \zeta (7,3,3)+28220
   \zeta (13)-880 \pi ^2 \zeta (11)-62032 \zeta (11) \cr 
   & -2880 \zeta (3) \zeta (9)-\frac{368
   \pi ^4 \zeta (9)}{3}+5760 \pi ^2 \zeta (9)+4640 \zeta (5) \zeta (7) \cr 
   & -1120 \zeta (3)^2
   \zeta (7)-\frac{128 \pi ^6 \zeta (7)}{27}+\frac{128 \pi ^4 \zeta (7)}{5}+2400 \zeta
   (3) \zeta (5)^2 \cr 
   & +1600 \zeta (5)^2+640 \zeta (3)^2 \zeta (5)-\frac{128 \pi ^6 \zeta
   (5)}{63} \, .  
\end{align}
It is certainly of great interest to evaluate these highly non-trivial higher-loop periods independently by other means. Not surprisingly,  all these periods contain many zeta values in a complicated fashion.  However, as we have emphasised, all these complications cancel out leaving only $\zeta(13)$ in the final integrated correlators, as shown in \eqref{eq:master-In}. 


%

\end{document}